# Polarity-tunable magnetic tunnel junctions based on ferromagnetism at oxide heterointerfaces

**Authors:** Thach D. N. Ngo[1,2], Jung-Won Chang[1,3], Kyujoon Lee[4], Seungju Han[5], Joon Sung Lee[3], Young Heon Kim[1], Myung-Hwa Jung[4], Yong-Joo Doh[3], Mahn-Soo Choi[5], Jonghyun Song[6], Jinhee Kim[1]

**Affiliations:**

[1]Korea Research Institute of Standards and Science, Daejeon 305-340, Korea.

[2]Department of Nanoscience, Korea University of Science and Technology, Daejeon 305-350, Korea.

[3]Department of Applied Physics, Korea University Sejong Campus, Sejong 339-700, Korea.

[4]Department of Physics, Sogang University, Seoul 121-742, Korea.

[5]Department of Physics, Korea University, Seoul 136-713, Korea.

[6]Department of Physics, Chungnam National University, Daejeon 305-764, Korea.

Correspondence and requests for materials should be addressed to J.K. (e-mail: jinhee@kriss.re.kr) or to J.S. (email: songjonghyun@cnu.ac.kr).

Complex oxide systems have attracted considerable attention because of their fascinating properties, including the magnetic ordering at the conducting interface between two band insulators, such as $LaAlO_3$ (LAO) and $SrTiO_3$ (STO). However, the manipulation of the spin degree of freedom at the LAO/STO heterointerface has remained elusive. Here, we have fabricated hybrid magnetic tunnel junctions consisting of Co and LAO/STO ferromagnets with the insertion of a Ti layer in between, which clearly exhibit magnetic switching and the tunnelling magnetoresistance (TMR) effect below 10 K. The magnitude and the of the TMR are strongly dependent on the direction of the rotational magnetic field parallel to the LAO/STO plane, which is attributed to a strong Rashba-type spin orbit coupling in the LAO/STO heterostructure. Our study provides a further support for the existence of the macroscopic ferromagnetism at LAO/STO heterointerfaces and opens a novel route to realize interfacial spintronics devices.

Transition-metal oxide heterostructures with strongly correlated electrons have demonstrated a host of novel physical phenomena that do not exist in their constituent materials[1-3], and therefore, they can be considered as a playground for both electronics and spintronics. The quantum correlation resulting from strong electron interactions leads to the interplay of multiple degrees of freedom, such as spin, charge, lattice and orbital. One example of these systems is the heterointerface of $LaAlO_3$ (LAO) and $SrTiO_3$ (STO, ref. 4), where both constituents are non-magnetic band insulators, but the electronic reconstruction[5] at their interface completely changes the physical properties of the entire system. The emergence of unexpected behaviours, such as the creation of a high-mobility two-dimensional electron gas[4], superconductivity[6,7], ferromagnetism[8-11] and the coexistence of superconductivity and ferromagnetism[12,13], has been reported and has inspired many intense studies.

Since the first detection of the hysteretic magnetoresistance (MR) in the LAO/STO system[14] in 2007, much effort has been devoted to investigating the magnetism at this oxide interface. The existence of the ferromagnetic ordering was proved by magnetization and magnetotransport measurements[8] and torque magnetometry[12]. The sparsely distributed ferromagnetic patches were mapped using the scanning superconducting quantum interference device (SQUID) technique[9,10]. The mechanism of such interface ferromagnetism, nevertheless, remains a controversial issue[15,16] and the manipulation of the spin degree of freedom also presents an interesting challenge. Indeed, the LAO/STO heterostructures can be considered a promising framework for spin-tunnelling applications in which the oxide interface performs as a ferromagnet and the LAO layer serves as an insulating barrier.

## Results

**Device growth and characterizations.** For this work, we fabricate hybrid magnetic tunnel junctions (MTJs) wherein an LAO/STO interface is one of the two ferromagnets and the other is the conventional ferromagnetic metal Co. The intervening LAO thin film constitutes a barrier for the spin-polarized tunnelling. The LAO/STO heterostructures are prepared using a pulsed laser deposition process, while the Co top electrode with an Au capping layer and the back gate Au electrode are deposited by sputtering (see Methods). Hall measurement yields carrier density of $n$ = $2 \times 10^{13}$ cm$^{-2}$ and Hall mobility of $\mu = 10^{3}$ cm$^{2}$/V·s for the LAO/STO interface at zero gate voltage[17]. Fig. 1a presents a representative high-resolution transmission electron microscopy (HR-TEM) cross-sectional view of the device along the [001]$_{pc}$ direction, where the subscript "pc" designates the pseudocubic structure of the LAO thin film. The image shows that the thin LAO film is completely strained[18,19], and that the interface between the LAO film and STO substrate is atomically flat, abrupt and free from misfit dislocations (see Supplementary Information).

When the magnetic field is applied parallel to the substrate as shown in Fig. 1b, the tunnelling magnetoresistance (TMR) depends on the parallel and antiparallel alignments of the magnetizations of two ferromagnets. The relative resistance change is given by TMR = ($R_{AP}$ – $R_{P}$)/$R_{P}$, where $R_{P}$ and $R_{AP}$ designate the tunnelling resistances when the magnetizations are parallel and antiparallel, respectively. It should be emphasized that the epitaxial growth of the thin Ti interlayer between Co and LAO layers is essential for observing clear TMR effects. The insertion of a thin Ti layer reduces the tunnelling resistance by at least three orders of magnitude compared with that for direct contact of Co and LAO layers, as shown in Fig. 1c, which enables an effective spin-polarized tunnelling. Because the nonmagnetic metal layer may reduce the polarization of the spin current from the Co ferromagnet, and thus the TMR ratios, the optimal thickness of the Ti layer is extremely critical. By varying the thickness of the Ti layer, the tunnelling resistance-area (RA) product to obtain an optimal TMR ratio was revealed to be 1.25 – 2.50 $\Omega cm^2$ with 2.5 nm-thick Ti layer. The relatively low value of RA product is attributed to an effective reduction of the barrier thickness due to an ionic built-in potential in LAO (ref. 20), incorporated with low work function of the metal ($\Phi = 5.0$ eV for Co and 4.33 eV for Ti, refs 20, 21).

**Observation of TMR.** Typical TMR data acquired at low temperatures are presented in Fig. 2a, where the magnetic field is applied parallel to the crystallographic axis with $\alpha = 0°$. Abrupt switching of the positive TMR occurs at the coercive forces of Co [$H_c$(Co) ~ 40 Oe] and of the ferromagnetic LAO/STO interface [$H_c$(LAO/STO) ~ 600 Oe] with respect to the magnetization alignments of the two ferromagnets. The hysteretic magnetization cycle of the LAO/STO heterostructure sample in Fig. 2c reveals that $H_c$(LAO/STO) is approximately 600 Oe, which is also similar to that reported by Ariando *et al* .(see ref. 8 and Fig. S3 in Supplementary Information). Our observations of the clear and very sharp switching behaviour of the TMR in the hybrid MTJs, which is the first to demonstrate the hybrid MTJ utilizing the LAO/STO interface to the best of our knowledge, indicates that the interfacial ferromagnetism of the LAO/STO heterostructure occurs on the macroscopic scale. It is noted that the TMR ratio is suppressed at negative voltages and is almost negligible at $V_g = -25$ V (Fig. 2b). Because the $V_g$ dependence of the TMR ratios is quite similar to that of the tunnelling conductance in the zero-magnetic-field state (Fig. 2d), the vanishing of the TMR at $V_g = -25$ V can be attributed to a depletion of the charge carriers at the LAO/STO interface. Furthermore, the TMR signals exhibit a slight dependence on the bias current in the measurement range of 50 nA to 1 μA (see Fig. S4 in Supplementary Information).

**Anisotropy of TMR.** It is noticeable that a drastic change of the TMR effect occurs with the variation of the angle between the in-plane magnetic field and the crystallographic axes of the STO substrate, as displayed in Fig. 3a. The magnitude of the TMR, including its sign and coercive force, varies progressively with the angle of the magnetic field. The angle-dependent TMR ratios obtained from two different devices, M1 and M2, reveals two-fold rotational symmetric behaviour as observed in Fig. 3b. Note that the TMR signs of the two devices are opposite. This out-of-phase behaviour will be discussed later. Another interesting feature is the angle-dependent coercive force, exhibiting a four-fold symmetric behaviour, as shown in Fig. 3c. The ratio of the maximum to minimum coercive forces reaches 2.8 (1.7) for M1 (M2). This finding indicates that the STO[100] and [010] crystallographic axes correspond to the magnetic easy axis for the interfacial ferromagnetism of the LAO/STO heterostructure, which is the energetically favourable direction for the spontaneous magnetization, while the diagonal STO[110] axis behaves as the hard axis. Thus, the four-fold symmetric coercive force of the LAO/STO interface resembles the pseudo-cubic symmetry of the STO substrate[22,23], while the two-fold symmetric TMR signal is caused by the uniaxial anisotropy of the electric conduction at the LAO/STO interface[24-26].

A strong correlation between the interfacial ferromagnetism and STO structure can also be observed in Fig. 4, which displays various TMR behaviours depending on successive thermal cycling. It is quite striking that the sign of the TMR is reversed from negative to positive (I-II and III-IV stages) or vice versa (II-III), after the sample is heated to a temperature higher than 150 K. The thermal operation below 50 K, however, does not affect the sign and magnitude of the TMR (IV-V). Here, we speculate that the tetragonal domain structure[25] of the STO substrate is responsible for the sign reversal of the TMR. Because the STO substrate undergoes a structural phase transition at approximately 105 K, which is driven by a rotation of the oxygen octahedrons about the crystallographic axis[27], and the interfacial conductivity of the LAO/STO heterostructures is enhanced along the antiphase domain boundaries separating tetragonal domains in STO (ref. 25), the thermal cycling above 150 K induces a drastic change in the domain structure resulting in diverse TMR behaviours. The sign difference of the TMR detected in devices M1 and M2 (Fig. 3b) can be explained in a similar manner. Note that the coercive force of the LAO/STO interface is not affected by the thermal cycling.

## Discussions

It has been reported that the sign reversal of the TMR in various MTJs can be caused by modulation of the resonant tunnelling process by the gate voltage[28,29], device structure[30] and the band structure in tunnel junctions[31,32]. In particular, the TMR sign reversal was reported in the MTJs with STO and LAO tunnel barrier. A dc bias voltage in the range of V was applied and the bias dependence reflected the structure of the density of states of the ferromagnetic layers[31,32]. In our work, unlikely, the an ac scheme of bias voltage up to 1 mV allows to measure the spin transport around the Fermi level and the band structure can be considered unchanged within that small range[33]. In addition, the sign change of the TMR, triggered by the rotation of the magnetic field and the thermal cycling, is unique in our hybrid MTJs and cannot be explained by the anisotropic magnetoresistance (AMR) observed in the LAO/STO heterointerface under high magnetic fields[34-36]. Moreover, our observation differs from the tunnelling anisotropic magnetoresistance[37] (TAMR), which also exhibits angle-dependent sign reversal, in the following aspects: (i) without a Co layer no hysteretic magnetoresistance is observed and (ii) the two resistance jumps match well with the coercive forces of the two constituting ferromagnetic layers.

To explain the sign-switching anisotropic TMR observed in the hybrid junctions of Co and LAO/STO heterostructure, here we propose a possible scenario based on a simplified model[38] considering a strong spin-orbit coupling in the LAO/STO heterostructure. Our model is summarized as follows. As a recent experiment[25] suggests that the electrical conduction in the LAO/STO interface occurs mainly along the narrow paths at the twin boundaries of the STO substrate, which are elongated along the crystallographic axis. Thus, we regard these conducting paths in the LAO/STO interface as one-dimensional (1D) quantum wires (QWs) and for simplicity they are considered parallel. When the direct conduction between the QWs is ignored at the lowest approximation, the hybrid MTJ can be regarded as a two-dimensional junction consisting of a ferromagnetic top electrode (Co), insulating layer (LAO) and 1D QW with both magnetic ordering and Rashba spin-orbit coupling[39,40] (see Fig. S5 in Supplementary Information). The resistance along the QW can be ignored for simplicity, because the MR occurs mainly at the tunnel junction between Co and QW. For the conventional MTJs, the positive TMR occurs due to the relative alignment of the spin polarization direction of the two ferromagnetic electrodes, satisfying the Julliere's model[41]. However, in our hybrid MTJ model, TMR is determined by the momentum mismatch and the spin overlap across the tunnel junction, which are governed by the quantization of transversal momentum in the QW and the relative angle between the Zeeman and Rashba fields. Numerical calculation[38] using typical parameters of the LAO/STO heterostructure[42-45], such as the Fermi energy $E_F$ = 40 meV, the Rashba spin-splitting energy $\Delta_{SO} = E_F/2$, the in-plane Zeeman field $\Delta_{Zeeman} = E_F/16$, and the thickness of the LAO/STO interface $d = 1$ nm, shows that the sign and magnitude of the TMR vary depending on the angle between the in-plane magnetic field and the Rashba field directions, revealing two-fold rotational symmetric $\cos(\alpha/2)$ behaviour. This theoretical prediction is quite consistent to our observations of sign-switching and two-fold anisotropic TMR as shown in Fig. 3b. Furthermore, the out-of-phase TMR data in Fig. 3b obtained from two different samples can be understood by randomly chosen twin boundaries of the respective STO substrates, which are presumed to be elongated along two different perpendicular directions. Thermal-history-dependent sign reversal of the TMR in Fig. 4 also can be explained in a similar manner considering the change of domain structure upon the thermal cycle[25]. It should be emphasized that our work of the hybrid MTJ made of the LAO/STO interface provides a unique system to study the interplay of the magnetic ordering and Rashba spin-orbit coupling in addition to the interfacial ferromagnetism of the LAO/STO heterostructure.

In summary, we have fabricated hybrid MTJs consisting of Co and LAO/STO heterostructures and observed clear switching of the TMR, which is a clear support for the existence of large-scale interfacial ferromagnetism. The TMR sign is strongly dependent on the in-plane rotation of the magnetic field and thermal cycling. The hybrid MTJ based on the interfacial ferromagnetism provides a new platform to develop metal oxide spintronics devices. Because of the emergence of fascinating properties, the LAO/STO heterostructures can be considered a promising framework for multifunctional device development, in which several degrees of freedom are simultaneously considered. Our achievement of the spin injection and detection schemes for the heterointerfaces opens a new path to investigating the interfacial ferromagnetism and superconductivity.

## Methods

**Sample preparation.** The oxide heterointerfaces were prepared by growing 10 u.c. of LAO on TiO$_2$-terminated STO(001) substrates using pulsed laser deposition at 750°C with an oxygen pressure of $10^{-5}$ Torr. The STO substrates were commercial ones with a nominal thickness of ~ 0.5 mm. After growth, the samples were annealed *in situ* in a high-pressure oxygen environment (500 mTorr) for 30 minutes; this pressure was maintained during the cooling process. Subsequently, top metal electrodes 500×500 μm$^2$ in size, which comprised a few nm of Ti and 40 nm of Co, were deposited on the LAO film surface using sputtering deposition with a shadow mask. All the electrodes were capped with 20 nm of Au. Another Au electrode was deposited on the back of the STO substrates for gate tuning measurements. The samples M1, M2 and M3 were identical with a ~ 2.5-nm-thick Ti interlayer. Gold wires were attached to the top electrode with Ag paste and Al wires were attached to the oxide interface by wedge bonding.

**Measurements**. The current bias technique was used to determine the tunnelling resistance. An ac bias voltage was applied to the top electrode while the oxide interface was connected to the ground, ensuring that the current flowed along one of crystallographic axes of the oxide structure. The devices were mounted on a rotatable sample holder that enabled a 360° in-plane rotation of the magnetic field. The direction of the magnetic field was defined by its angle $\alpha$ with respect to the direction along the STO[100] axis. The magnetoresistance (MR) measurements were mainly performed at low temperatures with tunnel junctions consisting of a ~2.5-nm-thick Ti layer.

## Acknowledgements

This research was supported by the Basic Science Research Program of the National Research Foundation of Korea (NRF), funded by the Ministry of Education, Science and Technology (Grants 2012R1A2A2A02013936 and 2012R1A2A1A01008027).

## Author Contributions

T.D.N.N, J.S. and J.K. designed this project. M.S.C. suggested the potential role of twin boundaries and thermal cycling experiments. T.D.N.N, J.W.C. and J.K. prepared the samples and conducted the transport experiments. Y.H.K performed the electron microscopy. K.L. and M.H.J. measured the magnetization. T.D.N.N, J.S.L., M.S.C., M.H.J. and J.K. analyzed the data. J.H. and M.S.C. performed the model calculations. T.D.N.N., J.S., M.S.C., Y.J.D. and J.K. wrote the manuscript. All the authors contributed to the discussion and provided feedback on the manuscript.

## Additional Information

**Supplementary Information** accompanies this paper at

http://www.nature.com/naturecommunications

**Competing financial interests:** The authors declare no competing financial interests.

**Reprints and permission** information is available online at

http://npg.nature.com/reprintsandpermissions/

**How to cite this article:**

**Figure 1 | LaAlO₃/SrTiO₃ interface-based magnetic tunnel junctions. a**, Cross-sectional transmission electron microscope micrograph of Co/Ti/LaAlO₃/SrTiO₃ junctions and the diffractograms of Ti and LaAlO₃. **b**, Measurement scheme of the devices. The angle $\alpha$ is the angle between the STO[100] and the in-plane magnetic field. **c**, Control of the tunnelling resistance of the fabricated junctions based on the thickness of the Ti interlayer. The forward bias was applied to the top electrodes.

**Figure 2 | Typical TMR data acquired at low temperatures with sample M1. a**, Dependence of the TMR on temperature (measured with a 100 nA bias current). The arrows indicate the sweeping direction of the magnetic field. The colour code is the same throughout the article. **b**, The TMR data in response to the gate voltage $V_g$ taken at 2 K. **c**, The magnetization cycle of the LAO/STO interface as a function of the in-plane magnetic field. **d**, The response of the TMR signals and the conductance ($\sigma$) to the backgate voltage. The conductance is normalized to that at $V_g = 30$ V. The negative backgate voltage suppresses the carrier density at the LAO/STO interface, causing the removal of the spin tunnelling signals.

**Figure 3 | Anisotropy of the spin tunnelling phenomena probed in LAO/STO-based tunnel junctions**. **a**, Evolution of the tunnelling magnetoresistance (TMR) curves of sample M1 with the angle $\alpha$ at 2 K. **b**, The TMR ratios of samples M1 and M2 exhibits a two-fold symmetry as a function of the angle $\alpha$. The solid curves are guides to the eye. **c**, The coercive force (H_C) of the LAO/STO interface exhibits a four-fold symmetry.

**Figure 4 | The TMR sign is randomly selected with thermal cycling above 150 K. a**, Thermal history of sample M3 placed at an angle $\alpha = 0°$ with the data acquired at the stages labelled I–V. **b**, The TMR is measured at 2 K after thermal cycling from room temperature (I), to 150 K (II), to

220 K (III), to 220 K (IV) and finally to 50 K (V).

Fig. 1

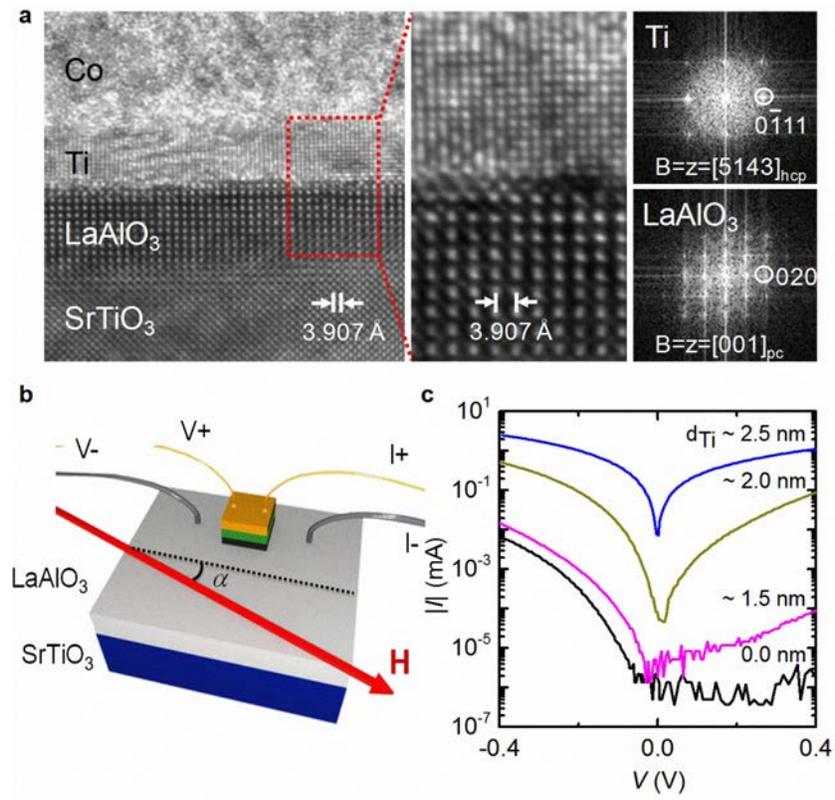

Fig. 2

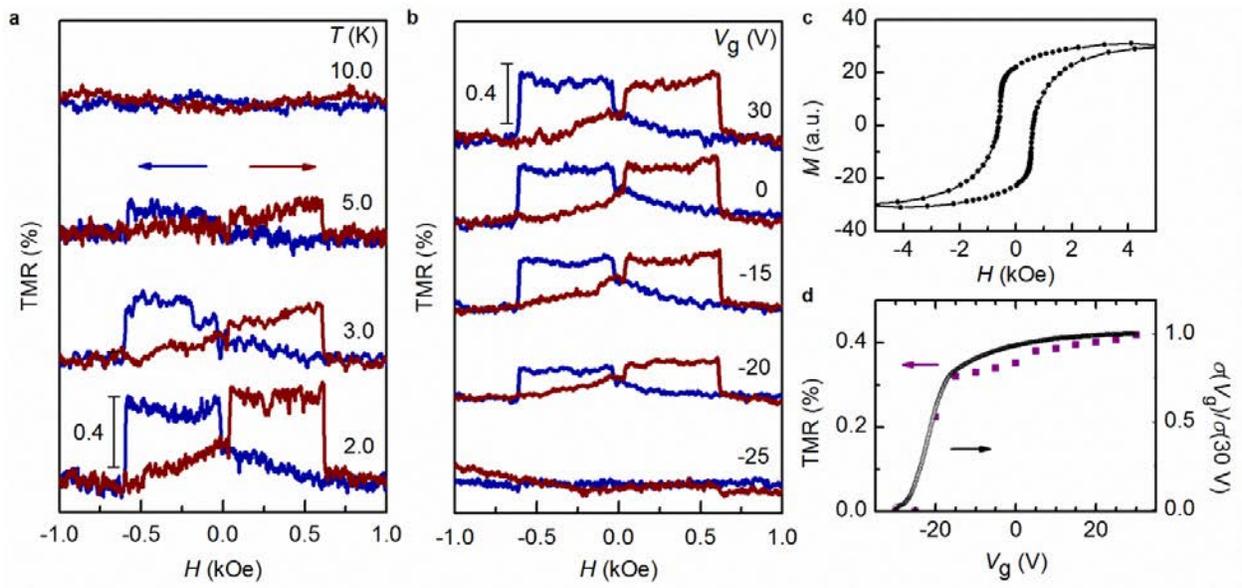

Fig. 3

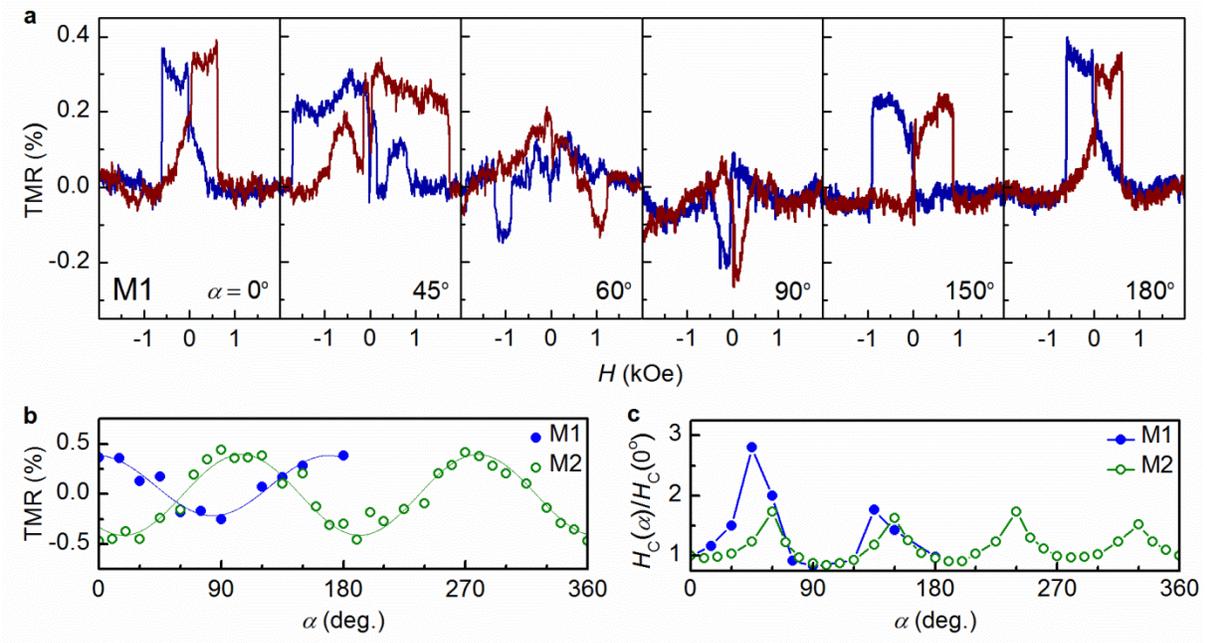

Fig. 4

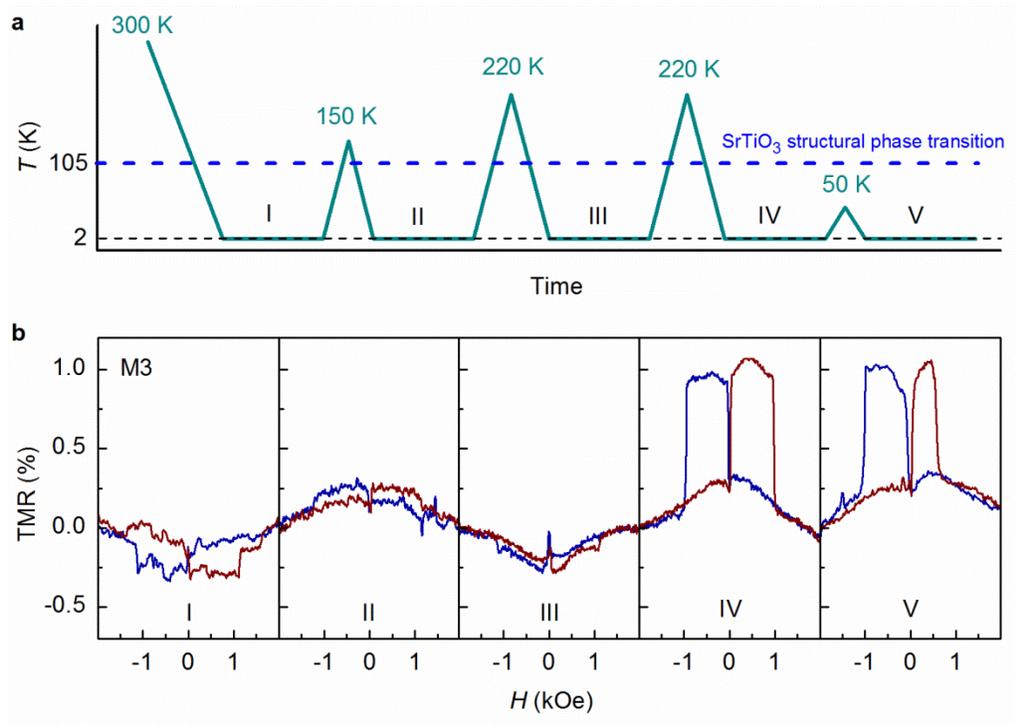

Fig. S1

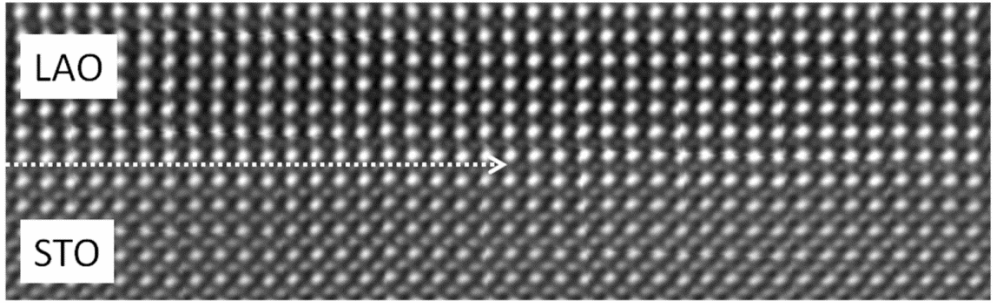

Fig. S2

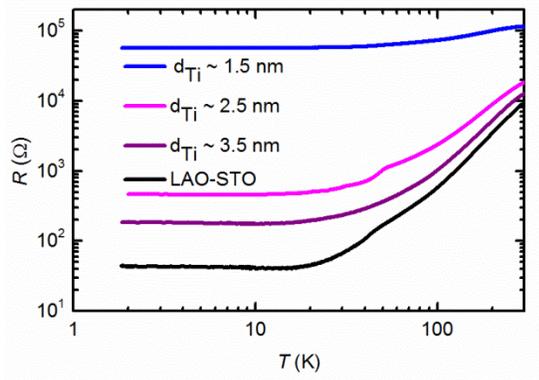

Fig. S3

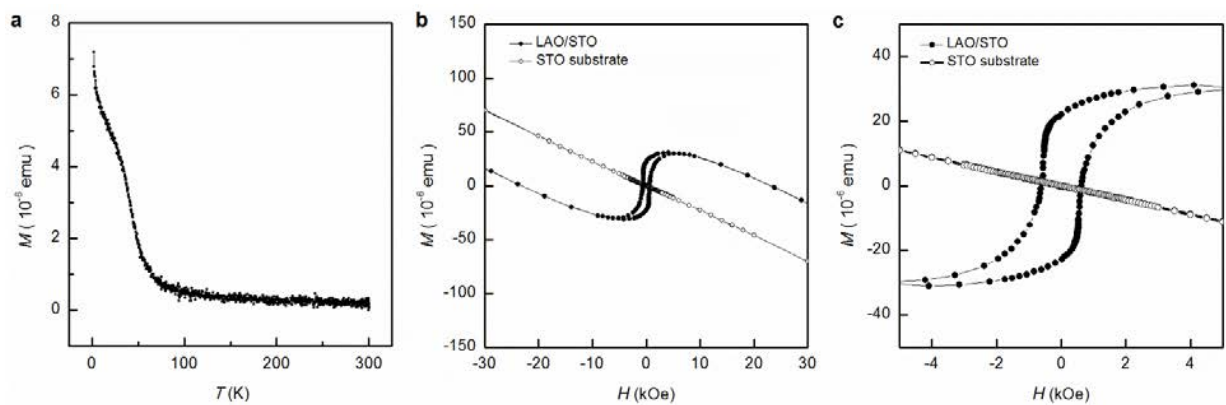

Fig. S4

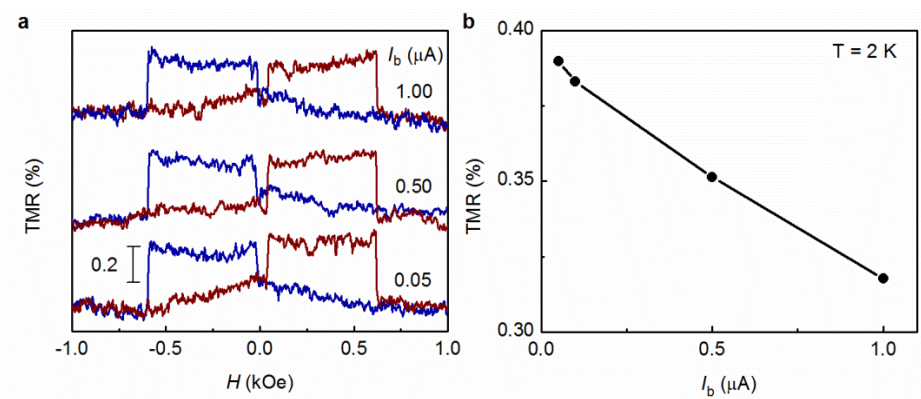

Fig. S5

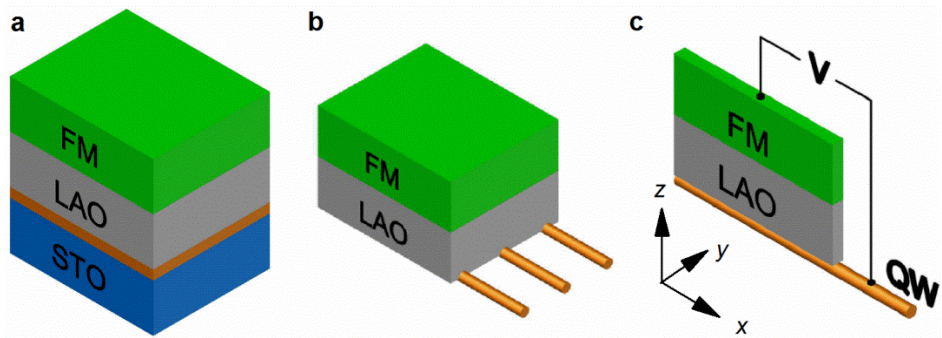